\documentclass[american,english,superscriptaddress,showkeys,amsmath]{revtex4}
\usepackage[T1]{fontenc}
\usepackage[latin9]{inputenc}
\usepackage{amstext}
\usepackage{amssymb}
\usepackage{graphicx}

\makeatletter

\providecommand{\tabularnewline}{\\}

\@ifundefined{textcolor}{}
{%
 \definecolor{BLACK}{gray}{0}
 \definecolor{WHITE}{gray}{1}
 \definecolor{RED}{rgb}{1,0,0}
 \definecolor{GREEN}{rgb}{0,1,0}
 \definecolor{BLUE}{rgb}{0,0,1}
 \definecolor{CYAN}{cmyk}{1,0,0,0}
 \definecolor{MAGENTA}{cmyk}{0,1,0,0}
 \definecolor{YELLOW}{cmyk}{0,0,1,0}
 }

\makeatother

\makeatother

\usepackage{babel}
\begin{document}

\title{Unit cell dependence of optical matrix elements in tight-binding
theory: The case of zigzag graphene nanoribbons}

\author{Kondayya Gundra}

\altaffiliation{Permanent Address: Theoretical Physics Division, Bhabha Atomic Research Centre, Mumbai 400085, INDIA}

\email{naiduk@barc.gov.in, shukla@iitb.ac.in}

\author{Alok Shukla}

\affiliation{Department of Physics, Indian Institute of Technology, Bombay, Mumbai
400076 INDIA}
\begin{abstract}
In the tight-binding theory, momentum matrix elements (MMEs) needed
to calculate the optical properties are normally computed using a
formulation based on the gradient of the Hamiltonian in the ${\bf k}$
space. We demonstrate the inadequacy of this formulation by considering
the case of zigzag graphene nanoribbons. We show that one obtains
wrong values of MMEs, in violation of the well-known selection rules,
if the unit cell chosen in the calculations does not incorporate the
symmetries of the bulk. This is in spite of the fact that the band
structure is insensitive to the choice of the unit cell. We substantiate
our results based on group-theoretic arguments. Our observations will
open an avenue for proper formulation of MMEs. 
\end{abstract}

\keywords{Tight binding theory, Momentum matrix elements, Optical properties,
Graphene nanoribbons}

\maketitle

\section{Introduction}

\label{sec:intro}

Tight-binding theory is one of the conceptually simplest, and widely
used theories of the electronic structure of molecules, clusters,
and solids. And indeed, the newly emerging field of graphene\cite{graphene}
and related materials\cite{graphene-nano,Zhang,Kane,Peres,Son} such
as graphene nanoribbons (GNRs)\cite{GNR,Rhim,Jung,Castro} has seen
widespread use of the tight-binding approach for calculation of their
electronic structure,\cite{graphene-tbind,Fujita,Ezawa,Onipko,Malysheva}
and optical properties.\cite{reichl} However, because of the unknown
nature of the basis set associated with the tight-binding approach,
the calculation of MMEs needed for computing the optical properties
has always been a matter of debate. Blount\cite{blount} argued that
the momentum operator for a periodic system can be represented as
\begin{equation}
{\bf p}=\frac{m_{0}}{\hbar}{\bf \nabla}_{{\bf k}}H({\bf k}),\label{eq:blount}
\end{equation}
where $m_{0}$ is the free electron mass, and ${\bf \nabla}_{{\bf k}}H({\bf k})$,
represents the gradient of the Hamiltonian in the ${\bf k}$ space.
This expression was used by Dresselhaus and Dresselhaus\cite{dress}
as well as Smith\cite{smith} to perform early calculations of the
optical properties of solids using the tight-binding approach. Based
upon generalized Hellmann-Feynman theorem, Lew Yan Voon and Ram-Mohan,\cite{ram-mohan}
argued that Eq. \ref{eq:blount} is indeed the correct representation
of the momentum operator for calculating optical properties. Adopting
a gauge-invariant approach within the tight-binding formalism, Graf
and Vogel\cite{graf-vogel} obtained results in agreement with the
work of Lew Yan Voon and Ram-Mohan,\cite{ram-mohan}. Cruz \emph{et
al.}\cite{cruz} also indicated that, Eq. \ref{eq:blount} leads to
the correct computation of optical matrix elements.\emph{ }However,
in a recent analysis, Pedersen \emph{et al.}\cite{pedersen} pointed
out that Eq. \ref{eq:blount} is incomplete, in that it does not contain
the contribution of intra-atomic matrix elements. According to Pedersen
\emph{et al.},\cite{pedersen} the MMEs $\langle c({\bf k})|\mathbf{{\bf \textbf{{\bf p}}}}|v({\bf k})\rangle$
between the valence band states ($|v({\bf k})\rangle$) and the conduction
band states ($|c({\bf k})\rangle$) is given by

\begin{eqnarray}
\langle c(\mathbf{k})|{\bf {\bf p}}|v(\mathbf{k})\rangle & = & \frac{m_{0}}{\hbar}\sum_{\alpha\,\beta}C_{c\beta}^{*}(\mathbf{k})C_{v\alpha}(\mathbf{k})\mathbf{\nabla}_{\mathbf{k}}\langle\beta,\mathbf{k}|H|\alpha,\mathbf{k}\rangle\nonumber \\
 &  & +\frac{im_{0}(E_{c,\mathbf{k}}-E_{v,\mathbf{k}})}{\hbar}\sum_{\alpha\,\beta}C_{c\beta}^{*}(\mathbf{k})C_{v\alpha}(\mathbf{k})\mathbf{d_{\beta\alpha}},\label{eq:p-matel}
\end{eqnarray}
where, the valence band eigen state $|v(\mathbf{k})\rangle$ is expressed
as

\begin{equation}
|v(\mathbf{k})\rangle=\sum_{\alpha}C_{v\alpha}(\mathbf{k})|\alpha,\mathbf{k}\rangle;\;|\alpha,\mathbf{k}\rangle=\frac{1}{\sqrt{N}}\sum_{\mathbf{R}}e^{i\mathbf{k.}\mathbf{R}}|\alpha,\mathbf{R}\rangle,\label{eq:v-state}
\end{equation}
and a similar expression holds for the conduction band states \foreignlanguage{american}{$|c(\mathbf{k})\rangle$}.
In the equations above, $\mathbf{R}$ denotes a lattice vector, $N$
is the total number of unit cells in the system, $|\alpha,\mathbf{R}\rangle$
is the $\alpha$-th atomic orbital located in the unit cell at position
$\mathbf{R}$, $E_{c,{\bf k}}$($E_{v,{\bf k}}$) is the energy eigen
value of the conduction (valence) band and $\mathbf{d}_{\beta\alpha}=\langle\beta,0|\mathbf{r}|\alpha,0\rangle$
is the matrix element of the position operator $\mathbf{r}$ with
respect to the reference unit cell. Note that the second term on the
right hand side (r.h.s.) of Eq. \ref{eq:p-matel}, called intra-atomic
contribution\cite{pedersen}, is the extra term as compared to Eq.
\ref{eq:blount}. Sandu\cite{sandu} further examined the issue and
essentially agreed with the analysis of Pedersen \emph{et al.}\cite{pedersen}

Recently, while developing a correlated electron approach for computing
optical properties of GNRs\cite{kondayya}, we discovered that the
optical matrix elements of zigzag GNRs (ZGNRs) computed using Eqs.
\ref{eq:blount} or \ref{eq:p-matel} were crucially depend on the
choice of the unit cell. While the band structures of the ZGNRs in
question, as expected, were found to be independent of the nature
of the unit cell, however, correct values of optical matrix elements
were not obtained unless the chosen unit cell also incorporated the
point-group symmetry of the bulk. This result is counter-intuitive,
and surprising, because normally we believe that the computed physical
quantities for bulk systems should be independent of the choice of
the unit cell. This issue is particularly important for the case of
ZGNRs for which there is a certain ambiguity in the choice of unit
cells, as compared to the case of armchair GNRs (AGNRs) for which
a unique choice of the unit cell exists. To the best of our knowledge,
this unit cell dependence of the formalism based on Eqs. \ref{eq:blount}
or \ref{eq:p-matel}, has not been reported earlier, therefore, here
we aim to elaborate our findings, and to analyze our results, based
upon group theoretic arguments.

Remainder of this paper is organized as follows. In the next section
we discuss the theoretical aspects of this work. In particular, we
analyse the nature of unit cells chosen for the calculations from
the point of view of their point group symmetries, and, based on their
irreducible representations, deduce the optical selection rules. In
section \ref{sec:results} we present and analyse our results. In
particular, we find that the results obtained are fully consistent
with the optical selection rules deduced in section \ref{sec:theory}.
Furthermore, we also support our arguments by means of finite cluster
calculations. Finally, in section \ref{sec:conclusions} we present
our conclusions.

\section{Theory}

\label{sec:theory}

In the present work, we consider a nearest-neighbor tight-binding
(TB) model for GNRs, with zero site energies
\begin{equation}
H=\sum_{\left\langle ij\right\rangle ,\sigma}t_{i,j}(c_{i\sigma}^{\dagger}c_{j\sigma}+c_{j\sigma}^{\dagger}c_{i\sigma}),\label{eq:tbind}
\end{equation}
where $\left\langle ij\right\rangle $ implies nearest neighbors (NN),
$c_{i\sigma}^{\dagger}$ creates an electron of spin $\sigma$ on
the $p_{z}$ orbital of carbon atom $i$ (assuming that the ribbon
lies in the $xy$-plane, with the $x$-axis being the periodicity
direction), and $t_{ij}$ is the corresponding hopping matrix element.
In order to obtain the band structure and the corresponding Bloch
orbitals, the TB Hamiltonian of Eq. \ref{eq:tbind} is Fourier transformed,
and the corresponding matrix elements for the one dimensional (1D)
system under consideration are obtained as
\begin{equation}
H_{i,j}(k)=\sum_{n=-\infty}^{n=+\infty}e^{ikna}t_{i(na),j(0)}\label{eq:hkspace}
\end{equation}
where $i(na)$ represents the $i-$th orbital of the unit cell located
at position $na$, $n$ being an integer and $a$ is the lattice constant,
$j(0)$ represents the $j-$th orbital of the reference unit cell,
and $t_{i(na),j(0)}$ is the corresponding hopping element which is
non-zero only for the NN sites. The Hamiltonian obtained from Eq.
\ref{eq:hkspace} is diagonalized at different $k-$points to obtain
the band structure and the corresponding Bloch orbitals in the 1D
Brillouin zone. In order to compute the optical absorption spectrum
within the TB model, the MMEs $\langle c({\bf k})|\mathbf{{\bf \textbf{{\bf p}}}}|v({\bf k})\rangle$
need to be computed. For the purpose, we have used the formula proposed
by Pedersen\emph{ et al.}\cite{pedersen} (\emph{cf}. Eq. \ref{eq:p-matel}),
as against the original approach of Blount\cite{blount} embodied
in Eq. \ref{eq:blount}. In order to ensure the correctness of the
approaches, we performed calculations of the matrix element $\langle c({\bf k})|p_{x}|v({\bf k})\rangle$needed
to compute the absorption spectrum for the light polarized along the
$x-$direction, using both Eqs. \ref{eq:blount}, and \ref{eq:p-matel},
and found only quantitative differences. However, Eq. \ref{eq:p-matel}
is more general, and can also be used to compute the matrix element
$\langle c({\bf k})|p_{y}|v({\bf k})\rangle$ required for calculating
the absorption spectrum for the $y$-polarized light, by setting the
first term on its r.h.s. to zero, because for a 1D system periodic
along the $x$ direction, the Hamiltonian has no $k_{y}$ dependence.

For the present 1D systems, we used the optical matrix elements to
compute the optical absorption spectrum for the $x$-polarized ($y$-polarized)
photons in form of the corresponding components of the imaginary part
of the dielectric constant tensor, i.e., $\epsilon_{xx}^{(2)}$($\epsilon_{yy}^{(2)}(\omega)$),
using the standard formula
\begin{equation}
\epsilon_{ii}^{(2)}(\omega)=C\sum_{v,c}\int_{-\pi/a}^{\pi/a}\frac{|\langle c(k)|p_{i}|v(k)\rangle|^{2}}{\{(E_{cv}(k)-\hbar\omega)^{2}+\gamma^{2}\}E_{cv}^{2}(k)}dk,\label{eq:eps2}
\end{equation}
where $i$ denotes the Cartesian direction in question, $\omega$
represents the angular frequency of the incident radiation, $E_{cv}(k)=E_{c,k}-E_{v,k}$,
$\gamma$ is the line width, while $C$ includes rest of the constants.
We have set $C=1$ in all the cases to obtain the absorption spectra
in arbitrary units. 

\begin{figure}
\includegraphics[width=7cm]{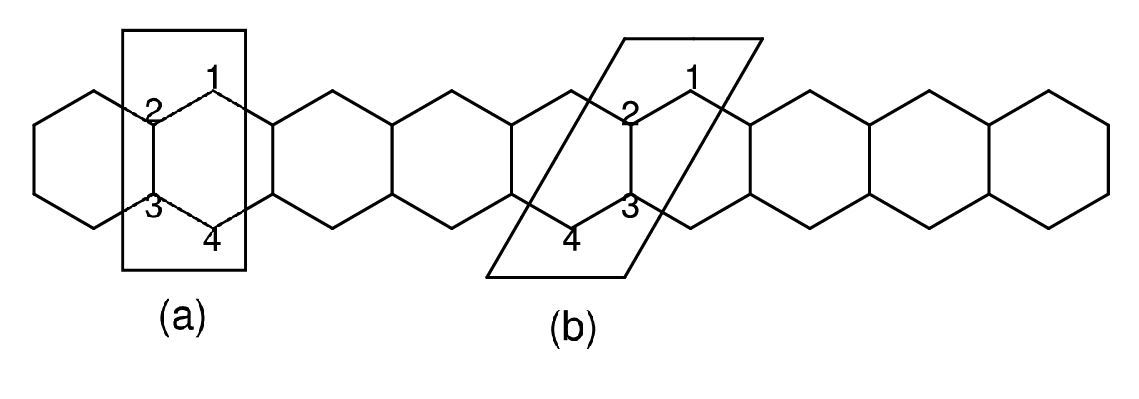}

\caption{ZGNR-2, along with the two possible choices of unit cells which can
generate it. The $x$-axis is taken along the periodicity direction,
$y-$axis is perpendicular to it within the plane of the system, while
the $z$-axis is perpendicular to that plane. Symmetry groups of cells
(a) and (b) are $C_{2v}$ and $C_{i}$, respectively. Based upon the
higher symmetry group of cell (a) we call it the symmetric cell, and
(b) the asymmetric cell. The numbering of the atoms is consistent
with the Hamiltonian matrices given in Eqs. \ref{eq:hmatsym} and
\ref{eq:hmatasym}.}

\label{Fig:cellZGNR}
\end{figure}

We will consider the optical matrix elements of various ZGNRs, characterized
by their width parameter $N_{Z}$, which is nothing but the number
of zigzag lines across the width of the ribbon. In short, we will
denote a ZGNR of width $N_{Z}$ as ZGNR-$N_{Z}$. We first consider
the narrowest such ribbon ZGNR-2, shown in Fig. \ref{Fig:cellZGNR},
which has four carbon atoms per unit cell and can be generated by
periodically repeating either of the two different unit cells depicted
in the same figure. Even though both types of unit cells lead to the
same GNR in the bulk limit, the point group symmetries of the two
unit cells are different. The unit cell of Fig. \ref{Fig:cellZGNR}a
has $C_{2v}$ symmetry with the symmetry operators (in addition to
the identity): (i) rotation by 180$^{\text{o}}$ about the $x$-axis,
(ii) reflection about the $xy$ plane, and (iii) reflection about
the $xz$ plane, leading to four irreducible representations (irreps)
$A_{1}$, $A_{2}$, $B_{1}$, and $B_{2}$.\cite{tinkham} As per
dipole selection rules of the $C_{2v}$ point group, $x$-polarized
radiation will couple bands with $A_{1}$ symmetry while the $y-$polarized
radiation will couple $A_{1}$ states to $B_{1}$ states.\cite{tinkham}
On the other hand, the unit cell of Fig. \ref{Fig:cellZGNR}b, has
$C_{i}$ symmetry with only the identity and the inversion being the
symmetry operators. This group has irreps $A_{g}$ and $A_{u}$, and
dipole selection rules allow $x$, $y$, and $z$, polarized (and
mixtures thereof) radiation to cause optical transitions between $A_{g}$
and $A_{u}$ bands. Thus, the symmetry analysis predicts that for
the $C_{2v}$ unit-cell (Fig. \ref{Fig:cellZGNR}a), the optical matrix
elements will have either $x$- or $y-$ non-zero components, while
for the $C{}_{i}$ case (Fig. \ref{Fig:cellZGNR}b)  both the components
will be simultaneously nonzero.

Using Eq. \ref{eq:hkspace}, and assuming that the NN hopping matrix
element is $t$, the Hamiltonian matrix for the symmetric unit cell
(\emph{cf}. Fig. \ref{Fig:cellZGNR}a) is obtained to be
\begin{equation}
H_{S}(k)=\left(\begin{array}{cccc}
0 & t(1+e^{-ika}) & 0 & 0\\
t(1+e^{ika}) & 0 & t & 0\\
0 & t & 0 & t(1+e^{ika})\\
0 & 0 & t(1+e^{-ika}) & 0
\end{array}\right),\label{eq:hmatsym}
\end{equation}
while for the asymmetric unit cell (Fig. \ref{Fig:cellZGNR}b) the
corresponding matrix is 
\begin{equation}
H_{AS}(k)=\left(\begin{array}{cccc}
0 & t(1+e^{-ika}) & 0 & 0\\
t(1+e^{ika}) & 0 & t & 0\\
0 & t & 0 & t(1+e^{-ika})\\
0 & 0 & t(1+e^{ika}) & 0
\end{array}\right).\label{eq:hmatasym}
\end{equation}

In order to compute the optical matrix elements (\emph{cf}. Eq. \ref{eq:p-matel}),
we need the matrix elements of the $\nabla_{{\bf k}}H({\bf k})$ operator,
which for the present 1D case, can be obtained easily by taking the
derivative of each element of $H_{S}(k)$/$H_{AS}(k)$, with respect
to $k$. Furthermore, the ${\bf d}$ matrix (\emph{cf}. Eq. \ref{eq:p-matel})
needed for the purpose is taken to be diagonal, with its elements
being the Cartesian coordinates of various atoms of the reference
unit cell.

Next, we demonstrate that that the Hamiltonian matrices of Eqs. \ref{eq:hmatsym}
and \ref{eq:hmatasym} are unitarily equivalent. Let us consider the
$k$-dependent unitary transformation 
\begin{equation}
U(k)=\left(\begin{array}{cccc}
1 & 0 & 0 & 0\\
0 & 1 & 0 & 0\\
0 & 0 & 1 & 0\\
0 & 0 & 0 & e^{ika}
\end{array}\right).\label{eq:unitary}
\end{equation}
It is straightforward to show that $H_{S}(k)=U(k)^{\dagger}H_{AS}(k)U(k)$.
Therefore, it is obvious that both $H_{S}(k)$ and $H_{AS}(k)$, will
yield the identical band structure.

\section{Results and Discussion}

\label{sec:results}

Here we present and discuss our numerical results in two subsections,
organized as follows: the first subsection contains the results for
various ribbons in the bulk limit (infinite length), while in the
second one the results are presented for finite-clusters of increasing
sizes for the two choices of the unit cell. As shown below, the finite-cluster
calculations help us in obtaining a better understanding of the results
for the infinitely long ribbons.

\subsection{Infinite Ribbons}

\label{sub:bulk-results}

\begin{figure}
\caption{Band structure of ZGNR-2 (\emph{cf}. Fig. \ref{Fig:cellZGNR}) obtained
by diagonalizing either $H_{S}(k)$ or $H_{AS}(k)$ (\emph{cf}. Eqs.
\ref{eq:hmatsym} and \ref{eq:hmatasym}) with $t=-2.6$ eV. Valence
bands are labeled as v$_{\text{1}}$/v$_{\text{2}}$ and conduction
bands as c$_{\text{1}}$/c$_{\text{2}}$.}

~\vspace*{\bigskipamount}

\includegraphics[width=6cm]{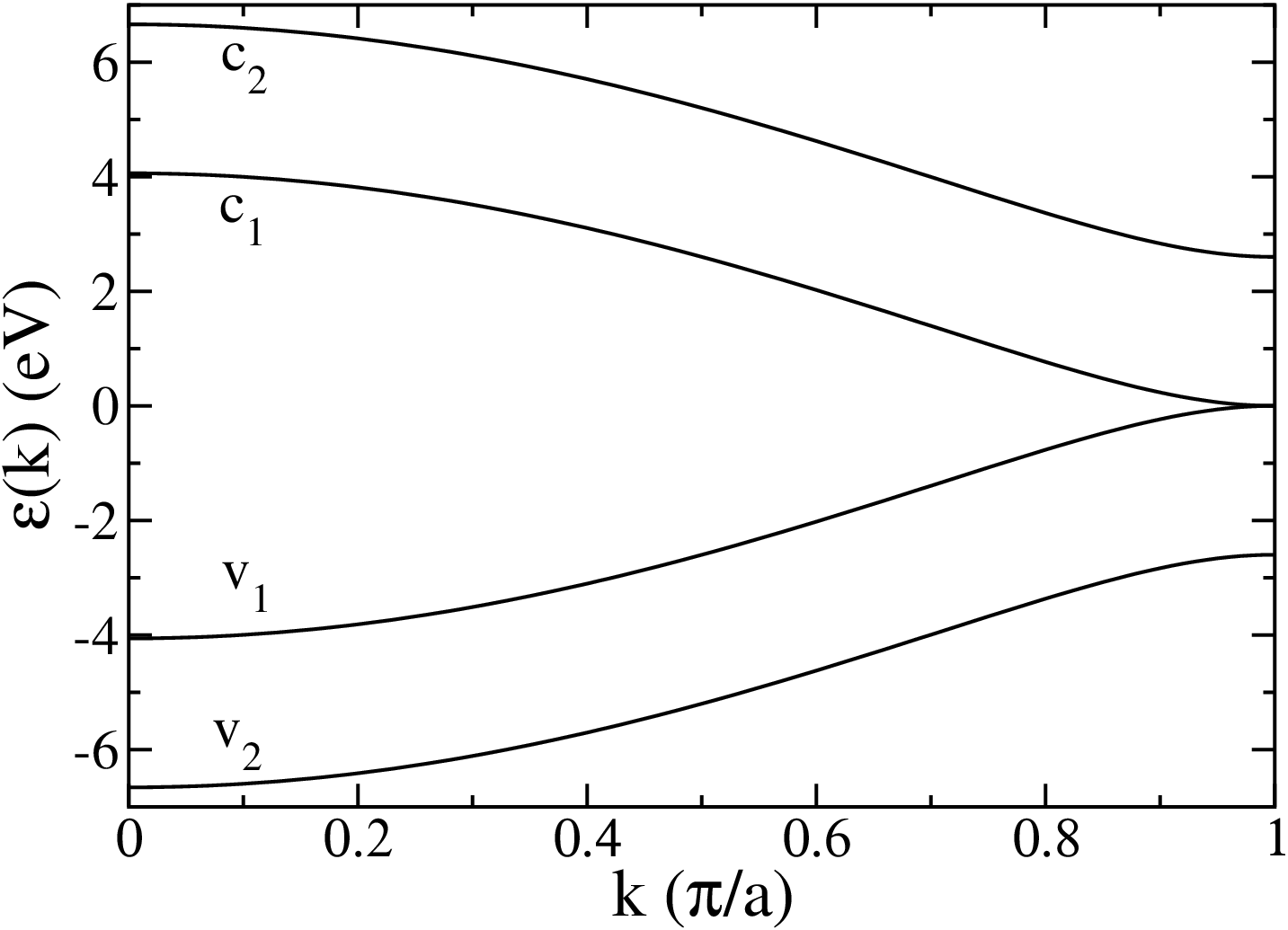}

\label{Fig:band-structure}
\end{figure}

The band structure of ZGNR-2 obtained by diagonalizing either $H_{S}(k)$
or $H_{AS}(k)$ is identical, and is presented in Fig. \ref{Fig:band-structure}\ 
for $t=-2.6$ eV. The irreps of the corresponding Bloch orbitals at
$k=0$ and $k=\pi/a$, for the two choices of the unit cell are listed
in Table \ref{tab:orb-irreps}, and can be easily deduced from the
Bloch orbitals given in the appendix.

\begin{table}
\caption{Irreps of the Bloch orbitals of ZGNR-2 at $k=0$ and $k=\pi/a$, for
the symmetric and the asymmetric choices of the unit cell. The Bloch
orbitals are presented in the appendix.}

\centering{}%
\begin{tabular}{|c||c|c|c|c|}
\hline 
Band & \multicolumn{4}{c|}{Irreps of the Bloch orbitals}\tabularnewline
\hline 
 & \multicolumn{2}{c|}{$k=0$} & \multicolumn{2}{c|}{$k=\pi/a$}\tabularnewline
\cline{2-5} 
\multicolumn{1}{|c|}{} & Sym ($C_{2v}$) & Asym ($C_{i}$) & Sym ($C_{2v}$) & Asym ($C_{i}$)\tabularnewline
\hline 
\hline 
$v_{2}$ & \multicolumn{1}{c||}{$A_{1}$} & $A_{g}$ & $A_{1}$ & $A_{g}$\tabularnewline
\hline 
\multicolumn{1}{|c|}{$v_{1}$} & $B_{1}$ & $A_{u}$ & $B_{1}$ & $A_{g}$\tabularnewline
\hline 
$c_{1}$ & \multicolumn{1}{c||}{$A_{1}$} & $A_{g}$ & $A_{1}$ & $A_{u}$\tabularnewline
\hline 
\multicolumn{1}{|c|}{$c_{2}$} & $B_{1}$ & $A_{u}$ & $B_{1}$ & $A_{u}$\tabularnewline
\hline 
\end{tabular}\label{tab:orb-irreps}
\end{table}

In Table \ref{tab:opt-mat} we present the values of optical matrix
elements between different bands for both the symmetric and asymmetric
unit cell cases, and for both types of unit cells the calculations
were performed with Eq. \ref{eq:p-matel}. It is a well-known fact
that in ZGNRs with even values of $N_{Z}$, and for light polarized
along the periodicity direction ($x$-axis), HOMO-LUMO transition
is forbidden, as are those between several other bands due to symmetry
related selection rules.\cite{reichl} Combining the knowledge of
the irreps of various Bloch orbitals (\emph{cf}. Table \ref{tab:orb-irreps})
at different $k$-points, with the dipole selection rules of point
groups $C_{2v}$ discussed in section \ref{sec:theory}, we note from
table \ref{tab:opt-mat} that, consistent with these selection rules,
the $x-$component of the optical matrix elements between bands $c_{1}-v_{1}$
(HOMO-LUMO) vanishes only if the symmetric unit cell is considered
for the system. From the same table it is also obvious this transition
is allowed for the $y-$polarized radiation when the symmetric cell
is used, consistent again with the selection rules of the $C_{2v}$
point group. While, with the asymmetric unit cell, both the $x$ and
$y$ components of the optical matrix element are found to be non-zero
for the transition, a result in agreement with the $C_{i}$ symmetry
of the asymmetric cell. Thus, we obtain different results for the
polarization characteristics of the radiation for the $c_{1}-v_{1}$
transition: (a) with the symmetric cell it is strictly $y$ polarized,
while (b) with the asymmetric cell it has both $x$ and $y$ components.
As far as the $c_{2}-v_{1}$ (and $c_{1}-v_{2}$) transitions are
concerned both the unit cells predict it to be $x$ polarized, however,
the magnitude of the optical matrix element obtained with the asymmetric
unit cell is much smaller as compared to that with the symmetric cell. 

\begin{figure}
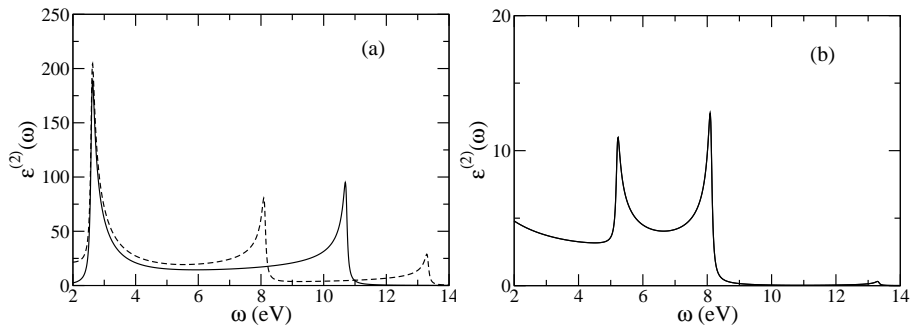

\caption{Components of the imaginary part of the frequency dependent dielectric
constant tensor of ZGNR-2, calculated using $t=-2.6$ eV: (a) $\epsilon_{xx}^{(2)}(\omega)$
for the symmetric unit cell (solid line), and asymmetric unit cell
(broken line), and (b) $\epsilon_{yy}^{(2)}(\omega)$ is identical
for both the unit cells. A line width of 0.05 eV was used to plot
the spectrum. \vspace{0.8cm}
}
\includegraphics[width=6cm]{fig3a}\includegraphics[width=6cm]{fig3b}\label{Fig:zgnr-2-spec}
\end{figure}

In Fig. \ref{Fig:zgnr-2-spec} we present the optical absorption spectrum
of ZGNR-2, for both the $x-$ and $y$-polarized radiation, calculated
for the two choices of the unit cell. From the band structure of ZGNR-2
(\emph{cf}. Fig. \ref{Fig:band-structure}) it is obvious that the
joint density of states (JDOS) of the conduction and valence bands
have van Hove (vH) singularities at points $k=0$ and $k=\pi/a$,
because of the parallel bands. This implies that, the optical absorption
spectrum for the $x$-polarized light, will have peaks corresponding
to gaps between those bands at points $k=0$ and $k=\pi/a$, for which
the transition is allowed by the selection rules. Therefore, we expect
two peaks in $\epsilon_{xx}^{(2)}(\omega)$ for ZGNR-2: (a) first
peak around 2.5 eV corresponding to $c_{1}-v_{2}$/$c_{2}-v_{1}$
allowed transitions at $k=\pi/a$, and (b) the second peak close to
11 eV because of the transition between the same bands at $k=0$.
For the $y$-polarized radiation, based upon the selection rules and
the vH singularities we expect three peaks in $\epsilon_{yy}^{(2)}(\omega)$:
(a) first one near 5 eV due to the $c_{2}-v_{2}$ transition at $k=\pi/a$,
(c) second one close to 8 eV due to the $c_{1}-v_{1}$ transition
at $k=0$, and (d) the final one near 13 eV caused by the $c_{2}-v_{2}$
transition also at $k=0$.

An inspection of Fig.\ref{Fig:zgnr-2-spec}b reveals that the calculated
values of $\epsilon_{yy}^{(2)}(\omega)$ exhibit precisely the three
peaks (the last peak near 13 eV being barely visible) described above,
both, for the symmetric, as well as for the asymmetric, unit cells.
The fact that the results for $\epsilon_{yy}^{(2)}(\omega)$ are same
with both choices of the unit cell is fairly obvious because the optical
matrix elements needed to compute $\epsilon_{yy}^{(2)}(\omega)$ depend
only on the $y$ coordinates of the sites (\emph{cf}. Eq. \ref{eq:p-matel}),
which are identical for both the unit cells. This is also obvious
from the $y$-components of the optical matrix elements listed for
various $k$-points in Table \ref{tab:opt-mat}. 

For $\epsilon_{xx}^{(2)}(\omega)$, however, the situation is different.
The calculated spectrum for the symmetric unit cell is fully consistent
with the vH singularity based analysis of the JDOS, with the two peaks
precisely at the predicted locations in Fig. \ref{Fig:zgnr-2-spec}a.
However, in the same figure, the spectrum computed with the asymmetric
unit cell agrees with this picture only for the first peak, while
it contains two higher energy peaks located near 8 eV and 13 eV, in
complete disagreement with the correct spectrum. From Table \ref{tab:opt-mat}
it is obvious that the optical matrix elements for both the symmetric
and asymmetric unit cells are identical at the point $k=\pi/a$, therefore,
even with the asymmetric unit cell we get the correct description
of the first peak in $\epsilon_{xx}^{(2)}(\omega)$. However, the
peaks around 8 eV and 13 eV in the asymmetric cell spectrum correspond
to $c_{1}-v_{1}$ and $c_{2}-v_{2}$ transitions, respectively, at
$k=0$, which have become allowed because of the incorrectly calculated
values of the corresponding optical matrix elements (\emph{cf.} Table
\ref{tab:opt-mat}). Thus, the peak around 8 eV is now present both
in $\epsilon_{xx}^{(2)}(\omega)$ and $\epsilon_{yy}^{(2)}(\omega)$
computed with the asymmetric cell, leading to absorption with mixed
polarization, a wrong result as discussed earlier. Furthermore, the
peak close to 11 eV is missing from the asymmetric cell spectrum because
for that case transitions $c_{1}-v_{2}$/$c_{2}-v_{1}$ have become
wrongly disallowed at $k=0$ (\emph{cf.} Table \ref{tab:opt-mat}).
Thus we conclude that with the asymmetric unit cell, for ZGNR-2 while
one obtains the correct description of the lowest peak in the absorption
spectrum, the predictions for the higher energy peaks in the spectrum
are completely wrong. 

\begin{table}
\caption{Optical matrix elements of ZGNR-2 between the bands identified in
Fig. \ref{Fig:band-structure} at various points in the Brillouin
zone, with $k$ expressed in the units of $\pi/a$. Column heading
{}``Sym'' corresponds to the values for the symmetric unit cell
(Fig. \ref{Fig:cellZGNR}a) and {}``Asym'' denotes the results obtained
for the asymmetric unit cell (Fig. \ref{Fig:cellZGNR}b). }
\begin{tabular}{|c|c|c|c|c|c|c|c|c|}
\hline 
$k$ ($\pi/a$) & \multicolumn{2}{c|}{$\langle c_{1}|p_{x}|v_{1}\rangle$} & \multicolumn{2}{c|}{$\langle c_{1}|p_{y}|v_{1}\rangle$} & \multicolumn{2}{c|}{$\langle c_{2}|p_{x}|v_{1}\rangle$} & \multicolumn{2}{c|}{$\langle c_{2}|p_{y}|v_{1}\rangle$}\tabularnewline
\hline 
\hline 
 & Sym & Asym & Sym & Asym & Sym & Asym & Sym & Asym\tabularnewline
\hline 
0.0 & 0.00  & 25.45  & 4.10 & 4.10 & 27.04 & 0.00  & 0.00 & 0.00\tabularnewline
\hline 
$0.2$ & 0.00  & 22.88 & 3.77 & 3.77 & 24.50  & 0.04  & 0.00 & 0.00\tabularnewline
\hline 
0.5 & 0.00  & 12.02  & 2.25 & 2.25 & 13.90 & 0.38  & 0.00 & 0.00\tabularnewline
\hline 
0.8 & 0.00  & 1.56  & 0.41 & 0.41 & 5.00 & 2.42  & 0.00 & 0.00\tabularnewline
\hline 
1.0 & 0.00  & 0.00  & 0.00 & 0.00 & 6.76  & 6.76  & 0.00 & 0.00\tabularnewline
\hline 
\end{tabular}\label{tab:opt-mat}
\end{table}

\begin{figure}
\includegraphics[width=6cm]{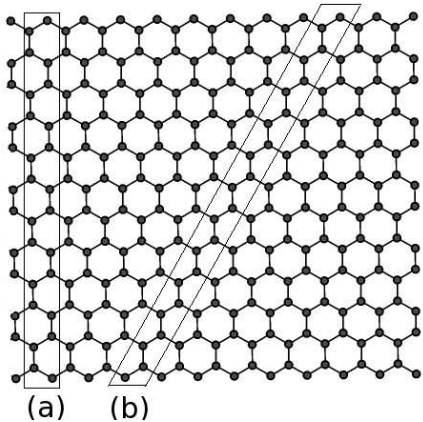}

\caption{Unit cells considered for the ZGNR-12: (a) symmetric unit cell, and
(b) asymmetric unit cell.\bigskip{}
}
\label{fig:cellBigZgnr}
\end{figure}

\begin{figure}
\includegraphics[width=6cm]{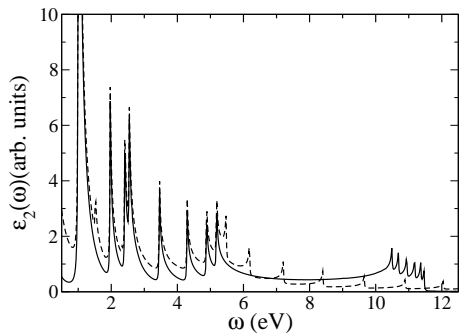}

\caption{ $\epsilon_{xx}^{(2)}(\omega)$ of ZGNR-12, computed using $t=-2.6$
eV, for the symmetric (solid line) and the asymmetric (broken line)
unit cells. A line width of 0.01 eV was used to plot the spectrum.}
\label{fig:opt-spectrum}
\end{figure}

In order to ensure that the results presented here for the ZGNR-2
are universal, and not just valid for this particular ZGNR, we also
performed similar calculations for the ZGNR-12. The symmetric and
asymmetric unit cells considered for this ZGNR are presented in Fig.
\ref{fig:cellBigZgnr}, and again the symmetric cell has $C_{2v}$
symmetry while the asymmetric one possesses $C_{i}$ symmetry. For
both types of unit cells we obtain the identical band structure (not
presented here), which is in agreement with that presented by other
authors.\cite{reichl} Because the number of bands, and, therefore,
the number of possible optical matrix elements is quite large for
this case, in Fig. \ref{fig:opt-spectrum} we directly present the
optical absorption spectra for the $x$-polarized photons ($\epsilon_{xx}^{(2)}(\omega)$)
of this ribbon computed using both types of unit cells. 

From the figure, a trend similar to the case of ZGNR-2 is observed
as far as the agreement between the spectra obtained using symmetric
and asymmetric unit cells are concerned. The two sets of spectra agree
at lower energy peaks, but diverge completely from each in the higher
energy region. As compared to the symmetric cell spectrum, the one
computed with the asymmetric cell in the high energy region has many
missing peaks as well as several new peaks. The reasons behind this
are precisely the same as for the case of ZGNR-2, that optical matrix
elements are being wrongly computed with the asymmetric cell, thus
allowing disallowed transitions and vice verse, leading to a spectrum
in complete disagreement with the published results.\cite{reichl}
In order to ascertain whether this behavior was an artifact of the
tight-binding model, we included electron-electron interactions by
employing the Pariser-Parr-People (PPP) model Hamiltonian\cite{PPP model}
to perform similar comparisons. Although the results of our PPP model
based work will be discussed in detail elsewhere,\cite{kondayya}but
we again found this unit-cell anomaly while computing the optical
matrix elements.

\subsection{Finite-Cluster Calculations}

\label{sub:finite-results}

\begin{figure}
\caption{Finite clusters containing $N$ repeat units of: (a) Symmetric cell,
and (b) Asymmetric cell }

\includegraphics[width=7cm]{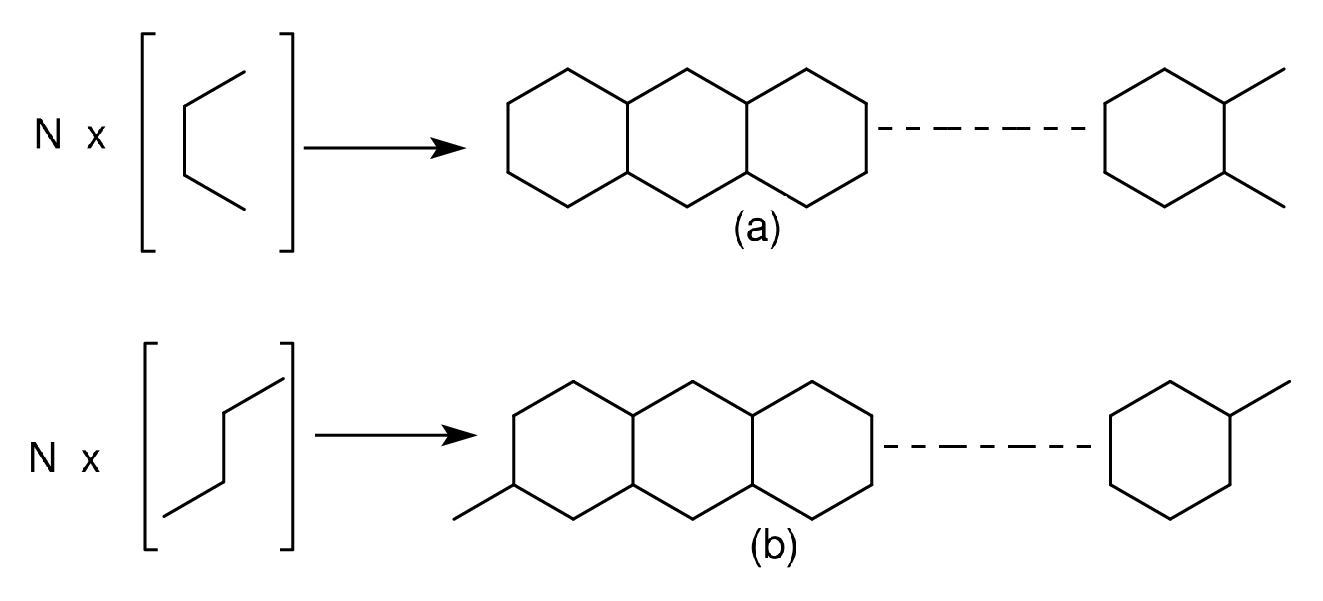}

\label{fig:finite-clusters}
\end{figure}

ZGNR-2, with hydrogenated edges is nothing but the polymer polyacene,
which has so far not been synthesized.\cite{acene-prb} However, several
oligomers of polyacene (oligoacenes) such as naphthalene, anthracene,
tetracene, pentacene, and hexacene are known to exist, and their properties
have been studied extensively both theoretically and experimentally.\cite{acene-prb}
The symmetry group of oligoacenes is $D_{2h}$, and the optical transitions
in it are either $x$- or $y$-polarized.\cite{acene-prb} The HOMO-LUMO
transition in oligoacenes is always $y$-polarized, just as in ZGNR-2,
while the dominant $x$-polarized transition is among the orbitals
HOMO(LUMO) and LUMO+$n$(HOMO-$n$), where $n$($\geq1$) is an integer
which depends on the oligoacene in question.\cite{acene-prb} Noting
that, the edge hydrogenation does not change the point group of the
system, optical selection rules of ZGNR-2 and polyacene will be identical.
Just as an oligoacene with an infinite number of repeat units is polyacene,
an infinite number of repeat units of either the symmetric cell or
the asymmetric cell will lead to ZGNR-2. Therefore, in what follows,
we investigate the optical properties of clusters of increasing sizes,
consisting of finite number of unit cells of the two types, to understand
their evolution towards the bulk ZGNR-2. The shape of the finite clusters
for the two cases is shown in Fig. \ref{fig:finite-clusters}. From
the figure it is obvious that, because of the dangling bonds at the
edges, the point groups of the clusters are $C_{2v}$ and $C_{i}$,
for the symmetric and the asymmetric cases, respectively. Furthermore,
without the dangling bonds, the symmetry group of these clusters will
be $D_{2h}$, just as in the case of oligoacenes. 

\begin{table}
\caption{HOMO-LUMO gap $E_{g}$ (in eV), $x$- and $y-$ components of dipole
transition matrix elements $\langle L|x|H\rangle$, and $\langle L|y|H\rangle$
between the HOMO and LUMO orbitals, respectively (in \AA\  units),
for finite clusters containing $N$ primitive cells of the symmetric
and asymmetric types.}

\begin{tabular}{|c|c|c|c|c|c|c|}
\hline 
$N$ & \multicolumn{3}{c|}{Symmetric Cell} & \multicolumn{3}{c|}{Asymmetric Cell}\tabularnewline
\hline 
 & $E_{g}$ & $\langle L|x|H\rangle/N$ & $\langle L|y|H\rangle$ & $E_{g}$ & $\langle L|x|H\rangle/N$ & $\langle L|y|H\rangle$\tabularnewline
\hline 
25 & 0.0197 & 0.000 & 1.392 & 0.0269 & 0.769 & 1.386\tabularnewline
\hline 
50 & 0.0050 & 0.000 & 1.398 & 0.0070 & 0.757 & 1.396\tabularnewline
\hline 
100 & 0.0013 & 0.000 & 1.400 & 0.0018 & 0.751 & 1.399\tabularnewline
\hline 
200 & 0.0003 & 0.000 & 1.399 & 0.0005 & 0.747 & 1.400\tabularnewline
\hline 
\end{tabular}\label{tab:finite-clus}
\end{table}

Table \ref{tab:finite-clus} contains our results on the HOMO-LUMO
gap, $E_{g}$, HOMO-LUMO dipole transition matrix elements $\langle L|x|H\rangle/N$\cite{dipole-x},
and $\langle L|y|H\rangle$ for the increasing cluster sizes. The
transition dipole moments were computed using the methodology adopted
in our earlier work.\cite{acene-prb} From the table it is obvious
that: (a) For both types of clusters, the gap $E_{g}$ is slowly closing
with increasing $N$, (b) $\langle L|y|H\rangle$ is virtually the
same for both types of clusters, and (c) $\langle L|x|H\rangle/N$
is exactly zero for all $N$ for the symmetric case, but it has significant
non-zero values for the asymmetric case, and demonstrates a slow decrease
with increasing $N$. This confirms the results of the previous section
that both types of unit cells yield the same band structure in the
bulk limit, but different values of the $x$ components of the transition
dipole moments.

\section{Conclusions}

\label{sec:conclusions}

In this work, we investigated the influence of the nature of the primitive
cells of ZGNRs on their electronic structure and optical properties,
as computed within the tight-binding model. Based upon the Bloch orbital
based calculations for the bulk systems, as well on finite clusters
of increasing sizes, we conclude that the choice of the unit cell
is of crucial importance when it comes to evaluation of optical matrix
elements within $k$-space gradient formulation, although, it is inconsequential
as far as the band structure is concerned. In other words, Hamiltonians
which are unitarily equivalent, lead to different values of optical
transition matrix elements. This result is surprising because the
theory of the optical properties of a system should not depend on
the choice of the unit cell adopted to describe it. Our work demonstrate
the inadequacy of this routinely used formulation for evaluation of
optical matrix elements. We further substantiated our findings based
upon the group theoretical analysis of optical matrix elements, and
conclude that a unit cell which does not have the symmetries of the
bulk system, leads to erroneous values of optical matrix elements
violating symmetry based selection rules. Therefore, in order to describe
the optical properties correctly within the $k$-space gradient formulation,
it is important to choose a unit-cell whose symmetry is consistent
with that of bulk. 

\emph{Acknowledgments: }We thank the Department of Science and Technology
(DST), Government of India, for providing financial support for this
work under Grant No. SR/S2/CMP-13/2006. K. G is grateful to Dr. S.
V. G. Menon (BARC) for his continued support of this work.

\appendix

\section{Bloch Orbitals of ZGNR-2 for the symmetric and asymmetric unit cells}

\label{sec:app-a}

The Bloch orbitals of ZGNR-2, obtained with the nearest neighbor hopping
matrix element of -2.6 eV is presented in table \ref{tab:orb-sym}
for symmetric cell and in table \ref{tab:orb-asym} for asymmetric
cell.

\begin{table}
\caption{Bloch orbital coefficients of ZGNR-2 at two $k$ points for the symmetric
cell. For each orbital, basis functions are numbered 1 to 4, from
top to bottom. For each coefficient, the real part is followed by
its imaginary part.}

\centering{}%
\begin{tabular}{|c|c|c|c|c|c|c|c|}
\hline 
\multicolumn{4}{|c|}{$k=0$} & \multicolumn{4}{c|}{$k=\pi/a$}\tabularnewline
\hline 
$v_{2}$ & $v_{1}$ & $c_{1}$ & $c_{2}$ & $v_{2}$ & $v_{1}$ & $c_{1}$ & $c_{2}$\tabularnewline
\hline 
\hline 
0.44, 0.00 & -0.56, 0.00 & 0.56, 0.00 & -0.44, 0.00 & 0.00, 0.00 & -0.71, 0.00 & 0.71, 0.00 & 0.00, 0.00\tabularnewline
\hline 
0.56, 0.00 & -0.44, 0.00 & -0.44, 0.00 & 0.56, 0.00 & 0.00, 0.71 & 0.00, 0.00 & 0.00, 0.00 & 0.00, 0.71\tabularnewline
\cline{1-1} \cline{3-8} 
0.56, 0.00 & 0.44, 0.00 & -0.44, 0.00 & -0.56, 0.00 & 0.00, 0.71 & 0.00, 0.00 & 0.00, 0.00 & 0.00, -0.71\tabularnewline
\hline 
0.44, 0.00 & 0.56, 0.00 & 0.56, 0.00 & 0.44, 0.00 & 0.00, 0.00 & 0.71, 0.00 & 0.71, 0.00 & 0.00, 0.00\tabularnewline
\hline 
\end{tabular}\label{tab:orb-sym}
\end{table}

\begin{table}
\caption{Bloch orbital coefficients of ZGNR-2 at two $k$ points for the asymmetric
cell. For each orbital, basis functions are numbered 1 to 4, from
top to bottom. For each coefficient, the real part is followed by
its imaginary part.}

\centering{}%
\begin{tabular}{|c|c|c|c|c|c|c|c|}
\hline 
\multicolumn{4}{|c|}{$k=0$} & \multicolumn{4}{c|}{$k=\pi/a$}\tabularnewline
\hline 
$v_{2}$ & $v_{1}$ & $c_{1}$ & $c_{2}$ & $v_{2}$ & $v_{1}$ & $c_{1}$ & $c_{2}$\tabularnewline
\hline 
\hline 
0.44, 0.00 & -0.56, 0.00 & 0.56, 0.00 & -0.44, 0.00 & 0.00, 0.00 & 0.71, 0.00 & -0.71, 0.00 & 0.00, 0.00\tabularnewline
\hline 
0.56, 0.00 & -0.44, 0.00 & -0.44, 0.00 & 0.56, 0.00 & 0.00, 0.71 & 0.00, 0.00 & 0.00, 0.00 & 0.00, 0.71\tabularnewline
\hline 
0.56, 0.00 & 0.44, 0.00 & -0.44, 0.00 & -0.56, 0.00 & 0.00, 0.71 & 0.00, 0.00 & 0.00, 0.00 & 0.00, -0.71\tabularnewline
\hline 
0.44, 0.00 & 0.56, 0.00 & 0.56, 0.00 & 0.44, 0.00 & 0.00, 0.00 & 0.71, 0.00 & 0.71, 0.00 & 0.00, 0.00\tabularnewline
\hline 
\end{tabular}\label{tab:orb-asym}
\end{table}


\begin{thebibliography}{References}
\bibitem{dipole-x}The $x$-component of the transition dipole was
divided by $N$, because it increases almost linearly with the increasing
cluster size. \pagebreak{}
\end{thebibliography}
\end{document}